# Reduced Bloch mode expansion for periodic media band structure calculations

Mahmoud I. Hussein

Department of Aerospace Engineering Sciences
University of Colorado at Boulder
Boulder, CO 80309-0429
Phone: (303) 472-3177
Fax: (303) 492-7881
Email: mih@colorado.edu

Number of pages of text:   15
Number of figures:          8

**Abstract**

Reduced Bloch mode expansion is presented for fast periodic media band structure calculations. The expansion employs a natural basis composed of a selected reduced set of Bloch eigenfunctions. The reduced basis is selected within the irreducible Brillouin zone at high symmetry points determined by the medium's crystal structure and group theory (and possibly at additional related points). At each of the reciprocal lattice selection points, a number of Bloch eigenfunctions are selected up to the frequency range of interest for the band structure calculations. Since it is common to initially discretize the periodic unit cell and solution field using some choice of basis, reduced Bloch mode expansion is practically a secondary expansion that uses a selected set of Bloch eigenvectors. Such expansion therefore keeps, and builds on, any favorable attributes a primary expansion approach might exhibit. Being in line with the well known concept of modal analysis, the proposed approach maintains accuracy while reducing the computation time by up to two orders of magnitudes or more depending on the size and extent of the calculations. Results are presented for phononic, photonic and electronic band structures.

## 1. INTRODUCTION

In the early years following the birth of quantum mechanics through the work of de Broglie, Schrödinger, Heisenburg and others[1], the nature of non-interacting electron motion in a perfect lattice was revealed by Felix Bloch through the formulation of *Bloch theory* [2]. His theory has rigorously, yet simply, formulated the concept of electron bands in crystals, hence the synonymous term *band theory*. Subsequent work based on band theory and the Pauli exclusion principle [3] has lead to the formal classification of all crystals into metals, semiconductors, and insulators [4–7]. Band theory has since played a major role in elucidating the electrical properties of crystalline solids. Band theory has also provided a basis for the study of optical, thermal and magnetic properties of crystals in association with the motion of other quantum particles: photons, phonons and magnons. In the late nineteen eighties, yet another type of periodic systems emerged only to ignite further research interest in Bloch wave propagation, namely, photonic and phononic crystals [8–10]. These are macroscale periodic materials that can be designed, via band engineering, to classically control the flow of light, sound and/or mechanical vibrations in a predetermined manner within the solid. In doing so, these modern materials have opened up a new technological frontier in optical, acoustic and elastic devices (see, for example, [11–14] and references within).

    Central to the application of band theory in all of the above disciplines is the process of band structure calculations. Due to crystallographic symmetry, the Bloch wave solution needs to be applied only to a single unit cell in the reciprocal lattice space covering the first Brillouin zone (BZ) [15]. Further utilization of symmetry reduces the solution domain, even more, to the irreducible Brillouin zone (IBZ). The target of the

---

[1] Brief history is given by Messiah [1], for example.

calculations is typically to generate a plot of the energy or frequency versus wave vector (along various directions inside or along the boundaries of the IBZ). The density of states, as a function of frequency, is also commonly computed. There are several techniques for band structure calculations. These include the planewave (PW) method [10,16–19], the transfer matrix method [20], the multiple scattering method [21,22], the finite difference method [23], the finite element (FE) method [24-26], the meshless method [27], the multiple multipole method [28], the wavelet method [29-30], the pseudospectral method [31], among others [32-34] (see [35] for a review). Some of the methods involve expanding the periodic domain (or potential) and the wave field using a truncated basis. This provides a means of classification in terms of the type of basis, e.g., the PW method involves a Fourier basis expansion and the FE method involves a real space basis expansion. The pros and cons of the various methods are discussed in depth in the literature, e.g. [35].

Regardless of the type of method used for band structure calculations, the computational effort involved is usually high because it involves solving a complex eigenvalue problem and doing so numerous times as the value of the wave vector, **k**, is varied. The size of the problem, and hence the computational load, is particularly high for the following cases. The first is when the unit cell configuration requires a large number of degrees of freedom to be adequately described. This, for example, could be due to a complex electronic potential or due to a complex unit cell material phase topology, both necessitating a finely resolved description. The second case is when the presence of defects is incorporated in the calculations. Defects are known to have a physical influence extending over very long ranges. This in turn requires choosing correspondingly large unit cells, known as supercells, for the band structure calculations. Consequently, large cells imply large numbers of degrees of freedom. The third case is concerned with an emerging area of research, that is, band structure optimization [36,37]. By varying certain parameters, the band structure can be tailored to desired form. Needless to say, optimization studies inherently involve numerous repeated forward calculations.

Some techniques have been developed to expedite band structure calculations, examples include utilization of the multigrid concept [38], development of fast iterative solvers for the Bloch eigenvalue problem [19, 24-25], and extension of homogenization methods to capture dispersion [39–43]. In this paper we provide a fundamentally different approach for fast band structure calculations. We present *reduced Bloch mode expansion*, which is an expansion employing a natural basis composed of a selected reduced set of Bloch eigenfunctions[2]. This reduced basis is selected within the IBZ at high symmetry points determined by the crystal structure and group theory (and possibly at additional related points). At each of these high symmetry points, a number of Bloch eigenfunctions are selected up to the frequency range of interest for the band structure calculations. As mentioned above, it is common to initially discretize the problem at hand using some choice of basis. In this manner, reduced Bloch mode expansion constitutes a secondary expansion using a set of Bloch eigenvectors, and hence keeps and builds on any favorable attributes a primary expansion approach might exhibit. The proposed approach is in line with the well known concept of modal analysis, which is widely used in various fields in the physical sciences and engineering. The concept of modal analysis is rooted in the idea of extracting a reduced set of representative information on the dynamical nature of a complex system. This practice is believed to have originated by the Egyptians in around 4700 B.C. in their quest to find effective ways to track the flooding of the Nile and predict celestial events. Over time spectral methods emerged as a tool for the study of cosmology and optics until modal expansion emerged, as presented by Rayleigh in his study of elastic vibrations, as a modeling paradigm in its own right [44]. For a thorough discussion on the historical development of modal analysis, the reader is referred to [45].

---

[2] The same mode selection concept, but in the context of a multiscale two-field variational method, was presented in [42].

In the next sections, a description of direct, full model-based calculation of band structures is given, followed by a description of the reduced Bloch mode expansion process and its application in a discrete setting using finite elements. Numerical examples are given for band structure, Bloch mode shape, and density of state calculations. Some results are then presented to demonstrate the efficiency of the calculations following the proposed approach. Linear elastic phononic media is chosen as a platform for presenting the formulations and results in these sections. Other sections cover the implementation of reduced Bloch mode expansion for photonic and electronic band structure calculations. Finally, a summary and conclusions section is provided.

## II. DIRECT IMPLEMENTATION OF BLOCH THEORY

Bloch theory describes the wavefunction of a particle in an infinite periodic medium, such as a crystal, in terms of wavefunctions at the reciprocal space vectors of a Bravais lattice[3]. The Bloch response, $\mathbf{f}$, is a product of a Bloch periodic function, $\tilde{\mathbf{f}}$, and a phase multiplier,

$$\mathbf{f}(\mathbf{x},\mathbf{k};t) = \tilde{\mathbf{f}}(\mathbf{x},\mathbf{k})e^{i(\mathbf{k}^T\mathbf{x}-\omega t)}, \qquad (1)$$

where $\mathbf{x} = \{x,y,z\}$ is the position vector, $\mathbf{k} = \{k_x, k_y, k_z\}$ is the wave vector, $i = \sqrt{-1}$, and $\omega$ and $t$ denote frequency and time, respectively. Bloch theory can also be expressed as

$$\mathbf{f}(\mathbf{x}+\mathbf{R},\mathbf{k}) = \mathbf{f}(\mathbf{x},\mathbf{k})e^{i(\mathbf{k}^T\mathbf{x})}, \qquad (2)$$

where $\mathbf{R}$ is the reciprocal lattice vector.

We now develop the application of Bloch theory in the context of three-dimensional (3D) phononic crystals modeled as a continuous linear elastodynamic periodic medium. Actual numerical implementation is covered in Section V A for a 2D plain strain model of a phononic crystal. The theory and implementation of Bloch theory for other types of periodic media, such as for photonic and electronic band structure calculations, are briefly presented in Sections V B and V C.

The governing continuum equation of motion for a heterogeneous medium is

$$\nabla \cdot \boldsymbol{\sigma} = \rho \ddot{\mathbf{u}}, \qquad (3)$$

---

[3] The 1D temporal version of Bloch theory is known as Floquet theory [46].

where $\boldsymbol{\sigma}$ is the stress tensor, $\mathbf{u}$ is the displacement field, $\rho$ is the density and a superposed dot denotes differentiation with respect to time. For an elastic medium,

$$\boldsymbol{\sigma} = \mathbf{C} : \nabla^S \mathbf{u}, \tag{4}$$

where $\mathbf{C}$ is the elasticity tensor and $\nabla^S$ denotes the symmetric gradient operator, that is,

$$\nabla^S \mathbf{u} = \frac{1}{2}\left(\nabla \mathbf{u} + (\nabla \mathbf{u})^T\right). \tag{5}$$

Substituting (4) into (3) yields

$$\nabla \cdot \mathbf{C} : \nabla^S \mathbf{u} = \rho \ddot{\mathbf{u}}, \tag{6}$$

which is the strong form of the general elastodynamic problem. Using operator notation, (6) can be expressed as

$$L[\mathbf{u}] = M[\ddot{\mathbf{u}}], \tag{7}$$

where $M[\mathbf{u}] = \rho \mathbf{u}$, and $L[\mathbf{u}] = \nabla \cdot \mathbf{C} : \nabla^S \mathbf{u}$. A 3D periodic medium is characterized by an infinitely repeated unit cell $\Omega$. In a phononic crystal, the unit cell is composed of two, or more, material phases. We will assume the material-to-material interfaces to be ideal. For the periodic unit cell domain $\Omega$, (6) has a Bloch solution of the form

$$\mathbf{u}(\mathbf{x}, \mathbf{k}; t) = \tilde{\mathbf{u}}(\mathbf{x}, \mathbf{k}) e^{i(\mathbf{k}^T \mathbf{x} - \omega t)}, \tag{8}$$

where $\tilde{\mathbf{u}}$ is the displacement Bloch function with the property $\tilde{\mathbf{u}} \in P_\Omega$, $P_\Omega = \{\tilde{\mathbf{u}} \mid \tilde{\mathbf{u}} \in H^1(\Omega), \tilde{\mathbf{u}}(\mathbf{x}) \text{ is } \Omega\text{-periodic}\}$. Using (8), the spatial component of the displacement gradient is

$$\nabla \mathbf{u} = \left(\nabla \tilde{\mathbf{u}} + i \mathbf{k}^T \otimes \tilde{\mathbf{u}}\right) e^{i \mathbf{k}^T \mathbf{y}}, \tag{9}$$

where the symbol $\otimes$ denotes outer product. Substitution of (5), (8) and (9) into (6) gives the strong form of the Bloch eigenvalue problem

$$\nabla \cdot \mathbf{C} : \left[\nabla^S \tilde{\mathbf{u}} + \frac{i}{2}\left(\mathbf{k}^T \otimes \tilde{\mathbf{u}} + \mathbf{k} \otimes \tilde{\mathbf{u}}^T\right)\right] = -\rho \omega^2 \tilde{\mathbf{u}}, \tag{10}$$

that is,

$$L^B[\tilde{\mathbf{u}}] = \lambda M[\tilde{\mathbf{u}}], \tag{11}$$

where $\lambda = -\omega^2$ and $L^B[\tilde{\mathbf{u}}] = \nabla \cdot \mathbf{C} : \left[ \nabla^S \tilde{\mathbf{u}} + \frac{i}{2}(\mathbf{k}^T \otimes \tilde{\mathbf{u}} + \mathbf{k} \otimes \tilde{\mathbf{u}}^T) \right]$ is the Bloch operator.

Using variational principles the weak form for (6) can be expressed as

$$\int_\Omega (\nabla^S \mathbf{w} : \mathbf{C} : \nabla^S \mathbf{u}) d\Omega = \int_Y (\rho \mathbf{w} \cdot \ddot{\mathbf{u}}) d\Omega, \tag{12}$$

where $\mathbf{w}$ is a weighting function. Upon Bloch transformation using (8), the weak form for the Bloch eigenvalue problem is [42]

$$\int_\Omega \overline{[\nabla^S \tilde{\mathbf{w}} + \frac{i}{2}(\mathbf{k}^T \otimes \tilde{\mathbf{w}} + \mathbf{k} \otimes \tilde{\mathbf{w}}^T)]} : \mathbf{C} : [\nabla^S \tilde{\mathbf{u}} + \frac{i}{2}(\mathbf{k}^T \otimes \tilde{\mathbf{u}} + \mathbf{k} \otimes \tilde{\mathbf{u}}^T)] d\Omega = -\omega^2 \int_Y \rho \overline{\tilde{\mathbf{w}}} \cdot \tilde{\mathbf{u}} d\Omega, \tag{13}$$

where $\tilde{\mathbf{w}}$ is the Bloch-transformed weighting function. The operator in (13) arising from the undergone transformation is Hermitian.

With periodic boundary conditions applied on $\Omega$, (13) is solved for wave vector $\mathbf{k}$ spanning the first Brillouin zone $\mathbf{k} \in [-\pi, \pi]^3$ over the unit cell. When symmetry allows (which is the usual case for crystals), it is sufficient for $\mathbf{k}$ to cover only the IBZ whose vertices are the high symmetry points determined by the crystal's structure and group theory (points shown in Fig. 1a). Following common practice, the $\mathbf{k}$-space for band diagram calculations is narrowed even further to only considering wave vectors pointing to positions along the border of this zone (circuit lines shown in Fig. 1). Considering a 2D square lattice, for example, the considered wave vector end points are along the paths $\Gamma \rightarrow X$, $X \rightarrow M$ and $M \rightarrow \Gamma$. For a 3D cubic lattice, the IBZ circuit consists of the paths $\Gamma \rightarrow X$, $X \rightarrow M$, $M \rightarrow R$ and $R \rightarrow \Gamma$. For a unit cell with side length $d$, $\mathbf{k} = [0, 0, 0]$ at $\Gamma$, $\mathbf{k} = [\pi/d, 0, 0]$ at X, $\mathbf{k} = [\pi/d, \pi/d, 0]$ at M and $\mathbf{k} = [\pi/d, \pi/d, \pi/d]$ at R. This process generates a representative band structure (frequency versus wave vector relation) for the periodic medium. For computational purposes, $\omega_p(\mathbf{k})$, where $p$ denotes branch number, is only evaluated at a discrete (uniformaly distributed) set of $n_k$ points along the entire IBZ border. The discrete wave vector set is denoted $\mathbf{k}_j = [k_1, k_2, k_3]_j$, $j = 1, \ldots, n_k$. The number of $\mathbf{k}$-points on a single straight line along the IBZ border connecting two high symmetry points is denoted $l_k$, i.e., $n_k = 3(l_k - 1) + 1$ and $n_k = 4(l_k - 1) + 1$ for the 2D and 3D cases, respectively, shown in Fig. 1. For density of states calculations, the $\mathbf{k}$-space sampling covers the entire IBZ, not only the border.

### III. REDUCED BLOCH MODE EXPANSION

In this section, the formulation for reduced Bloch mode expansion (RBME) is given as well as two eigenfunction selection schemes for the reduced basis.

### A. Formulation

Recall the Bloch eigenvalue problem given in (10) or (11). The displacement Bloch function is expressed as a superposition of Bloch mode eigenfunctions $\tilde{\mathbf{v}}_j(\mathbf{x})$, $j = 1, \ldots, \infty$,

$$\tilde{\mathbf{u}}(\mathbf{x}) = \sum_{j=1}^{\infty} \beta_j \tilde{\mathbf{v}}_j(\mathbf{x}), \qquad (14)$$

where $\beta_j$ is the coordinate for mode $j$. Substituting (14) into (11) gives

$$\sum_{j=1}^{\infty} \beta_j L^{\mathrm{B}}[\tilde{\mathbf{v}}_j] = \lambda \sum_{j=1}^{\infty} \beta_j M[\tilde{\mathbf{v}}_j]. \qquad (15)$$

Equation (15) is then multiplied by $\tilde{\mathbf{v}}_i(\mathbf{x})$, and integrated over the unit cell domain $\Omega$,

$$\sum_{j=1}^{\infty} \beta_j \int_{\Omega} ([\tilde{\mathbf{v}}_i] L^{\mathrm{B}}[\tilde{\mathbf{v}}_j]) d\Omega = \lambda \sum_{j=1}^{\infty} \beta_j \int_{\Omega} ([\tilde{\mathbf{v}}_i] M[\tilde{\mathbf{v}}_j]) d\Omega. \qquad (16)$$

If the Bloch eigenfunctions used in (14) correspond to the **k**-point eigenvalue problem for which the Bloch mode expansion is implemented, then this is essentially an orthogonal similarity transformation. Using the same eigenfunctions, or a broader selection of eigenfunctions as described in Section III B, for transforming the eigenvalue problem at other **k**-points is a deviation from this orthogonality condition. Furthermore, from a practical point of view only a truncated set of $q$ modes is retained for the Bloch mode expansion,

$$\tilde{\mathbf{u}}(\mathbf{x}) \cong \sum_{j=1}^{q} \beta_j \tilde{\mathbf{v}}(\mathbf{x}), \quad q \in \text{integer}, \qquad (17)$$

hence the targeted reduction in the size of the problem.

**B. Choice of Reduced Expansion Basis**

The proposed reduced Bloch mode expansion approach will only produce accurate results if a suitable choice of reduced basis is made. Here we propose two schemes for this selection. The first consists of choosing eigenfunctions corresponding to the high symmetry points that characterize the periodic lattice, as determined from crystallography and group theory. Fig. 1 shows these points as solid circles for a 2D square lattice and a 3D simple cubic lattice, i.e., the Γ, X, M points and the Γ, X, M, R points, respectively. The second scheme involves choosing eigenfunctions at all the high symmetry points mentioned above as well as at intermediate points centrally intersecting the straight lines joining these high symmetry points along the border of the IBZ. In practice, these would be points along the particular IBZ circuit lines that are intended for the band structure calculations, i.e., all the circles (solid and hollow) representing the Γ, Δ, X, Z, M, Σ points in 2D and the Γ, Δ, X, Z, M, T, R, Λ points in 3D, as shown in Fig. 1. This scheme is therefore richer and slightly more computationally expensive than the previous one. We refer to the first selection scheme as a *2-point expansion* scheme because eigenfunctions corresponding to 2 **k**-points on each of the straight lines bordering the IBZ are used. Similarly, we refer to the second selection scheme as a *3-point expansion* scheme. In principle, other schemes may be developed and can be tailored to the particular application and purpose for the band structure calculations. Another equally important consideration in setting up the reduced Bloch mode expansion basis is the number (and choice) of eigenfunctions retained at each of the **k**-points selected. In order for the band structure calculations to be accurate up to the frequency range of interest, this number should be equal to, or slightly higher than, the number of dispersion branches that are to be computed.

In this manner, both spatial and temporal dynamical characteristics are considered in the selection of the reduced Bloch mode expansion basis. The spatial characteristics are related to the selection of the **k**-points and the temporal characteristics are related to the choice of the eigenfunctions to retain at each of the selected **k**-points. Furthermore, this approach is similar in spirit to the concept of special **k**-points widely utilized in electronic structure analysis [47–48], as it involves using a reduced set of **k**-points for resourcefully carrying out broader calculations that involve information pertaining to the entire Brillouin zone. The utilization of symmetry has been used previously for other purposes, namely in the context of cutting the number of transfer operations in the transfer matrix method [49,50] and cutting the number of plane waves employed in the planewave method [51].

## IV. FINITE ELEMENT DISCRETIZATION

The finite element method is one of various methods for discretizing a periodic medium to facilitate numerical calculation of the band structure. A brief description of FE implementation for phononic band structure calculations is provided.

The Galerkin FE method (explained in detail in [52]) involves introducing a finite-dimensional approximation of the solution domain, its boundaries and the governing variational eigenvalue problem (13) that is to be solved. Upon discretization, the unit cell domain $\Omega$ is divided into $n_{el}$ finite elements,

$$\Omega = \bigcup_{e=1}^{n_{el}} \Omega^e, \tag{18}$$

where $\Omega^e$ denotes the domain for each element, $e = 1,\ldots, n_{el}$. The solution field, $\tilde{\mathbf{u}}(\mathbf{x})$, is discretized locally in domain $\Omega^e$ using well-chosen local piecewise polynomials, known as shape functions, $N_a$,

$$\tilde{\mathbf{u}}(\mathbf{x}) = \sum_a N_a(\mathbf{x})\mathbf{d}_a, \tag{19}$$

where $\mathbf{d}_a$ is the discrete displacement vector at node $a$. Upon application to (13) with incorporation of periodic boundary conditions, the following algebraic eigenvalue problem emerges

$$(\mathbf{K}(\mathbf{k}) - \omega^2 \mathbf{M})\tilde{\mathbf{U}} = \mathbf{0}, \tag{20}$$

where **M** and **K** are the global mass and stiffness matrices, respectively, and $\tilde{\mathbf{U}}$ is the discrete Bloch vector which is periodic in the domain $\Omega$. The global matrices **M** and **K** are assembled from the element-level matrices as shown in the Appendix. Also provided in the Appendix is the form of the finite element matrices associated with the Bloch transformation, i.e., the form of the *Bloch element*.

Solution of (20) at selected **k**-points provides the eigenvectors from which the reduced Bloch modal matrix, denoted $\boldsymbol{\Psi}$, is formed (as described in Section III). This matrix is used to expand $\widetilde{\mathbf{U}}$ in analogy to the continuous version of reduced Bloch mode expansion expressed in (17). In discrete form, the expansion follow the matrix multiplication

$$\widetilde{\mathbf{U}}_{(n \times 1)} = \boldsymbol{\Psi}_{(n \times m)} \widetilde{\mathbf{V}}_{(m \times 1)}, \tag{21}$$

where $\widetilde{\mathbf{V}}$ is a vector of modal coordinates for the unit cell Bloch mode shapes. In (21), $m$ and $n$ refer to the number of rows and number of columns for the matrix equation. For the **k**-point selection schemes described in Section III B, $m \ll n$. Substituting (21) into (20), and premultiplying by $\boldsymbol{\Psi}^{\mathrm{T}}$,

$$\boldsymbol{\Psi}^{\mathrm{T}} \mathbf{K}(\mathbf{k}) \boldsymbol{\Psi} \widetilde{\mathbf{V}} - \omega^2 \boldsymbol{\Psi}^{\mathrm{T}} \mathbf{M} \boldsymbol{\Psi} \widetilde{\mathbf{V}} = \mathbf{0}, \tag{22}$$

yields a reduced eigenvalue problem of size $m \times m$,

$$\overline{\mathbf{K}}(\mathbf{k}) \widetilde{\mathbf{V}} - \omega^2 \overline{\mathbf{M}} \widetilde{\mathbf{V}} = \mathbf{0}, \tag{23}$$

where $\overline{\mathbf{M}}$ and $\overline{\mathbf{K}}(\mathbf{k})$ are the generalized mass and stiffness matrices.

**Remark**

The fact that the operator associated with Bloch transformation is Hermitian is a favorable property for discretization and solution of the eigenvalue problem that emerges. The application of the reduced Bloch mode expansion approach requires for the Bloch eigenvectors to be orthogonal with respect the system matrices, e.g.., **M** and **K** in (20) for the elastodynamic problem. This in turn implies that the system matrices need to be Hermitian. While the Bloch operator is Hermitian, it is essential that this property is sustained through the discretization, such as the case with the FE method.

**V. NUMERICAL EXAMPLES**

**A. Phononic Band Structure Calculations**

We consider a linear elastic, isotropic, continuum model for a 2D phononic crystal under plain strain conditions. With this model, coupled in-plane P (longitudinal) and SV (shear vertical) wave propagation modes can be predicted. The algebraic Bloch eigenvalue problem is given in (20); details on the mass and stiffness matrices are provided in the Appendix. The **D** matrix for this model has the form

$$\mathbf{D} = \begin{bmatrix} \lambda + 2\mu & \lambda & 0 \\ \lambda & \lambda + 2\mu & 0 \\ 0 & 0 & \mu \end{bmatrix}, \quad (24)$$

where $\lambda$ and $\mu$ denote Lamé's constants.

As an example, a square lattice is considered with a bi-material unit cell. One material phase is chosen to be stiff and dense and the other compliant and light. In particular, a ratio of Young's moduli of $E_2/E_1 = 16$ and a ratio of densities of $\rho_2/\rho_1 = 8$ are chosen for all calculations presented. The topology of the material phase distribution in the unit cell is shown in Fig. 2. The unit cell is discretized into 45 × 45 uniformly sized 4-node bilinear quadrilateral finite elements, i.e., $n_{el} = 2025$. With the application of periodic boundary conditions, the number of degrees of freedom is $n = 4050$. The **k**-space is discretized such that $l_k = 49$, thus a total of $n_k = 145$ **k**-points are evaluated to generate the band structure. Fig. 3 shows the calculated band structure using 2-point expansion and 3-point expansion. Density of state calculations are also provided based on a total of $n_k = 561$ **k**-space sampling across the IBZ. In both sets of calculations, eight modes ($p = 1, \ldots, 8$) were selected at each of the selection points in **k**-space, i.e., a total of 24 eigenvectors ($m = 24$) were used to form the Bloch modal matrix for the 2-point expansion calculations. For the 3-point expansion calculations, $m = 48$. The results for the full model are overlaid for comparison indicating excellent agreement. The slightly more expensive 3-point expansion scheme gives better results at **k**-space regions far away from the high symmetry points Γ, X, M. The reduced Bloch mode expansion approach also performs well in predicting Bloch mode shapes as demonstrated in Fig. 4. The displacement field (based on 3-point expansion) in Fig. 4a and 4b represent, respectively, the Bloch mode shapes corresponding to the Γ point at the third branch (point A in Fig. 3b) and the Z point at the first branch (point B in Fig. 3b). Calculations using the same parameters were also carried out for internal segments in the IBZ, as illustrated in the insets of Fig. 5. The band structures (based on 2-point expansion) for the Γ→MZ, Γ→Z and Γ→ZX paths are shown in the figure. Excellent agreement is again observed in comparison with the full-model results. Finally, calculations for unit cells with elasticity contrast ratio of up to $E_2/E_1 = 2000$ (while keeping the density ratio the same) were also carried out (not shown) and the reduced model results accurately matched that of the full model.

**B. Photonic Band Structure Calculations**

The reduced Bloch mode expansion approach is now applied to photonic band structure calculations. A 2D model is considered for lossless electromagnetic waves propagating in-plane (i.e., the *x-y* plane). In this case, *TE* (**H** field in the *z* direction) and *TM* (**E** field in the *z* direction) polarized waves are, respectively, described by

$$\nabla \cdot \left( \frac{1}{\varepsilon_r(\mathbf{x})} \nabla H_z(\mathbf{x}) \right) + \frac{\omega^2}{c^2} H_z(\mathbf{x}) = 0, \quad (TE) \quad (25)$$

and

$$\nabla^2 E_z(\mathbf{x}) + \frac{\omega^2 \varepsilon_r(\mathbf{x})}{c^2} E_z(\mathbf{x}) = 0, \quad (TM) \tag{26}$$

where $\varepsilon_r$ is the dielectric constant and $c$ is the speed of light. The photonic crystal medium is periodic, i.e., $\varepsilon_r(\mathbf{x} + \mathbf{R}_{xy}) = \varepsilon_r(\mathbf{x})$, where $\mathbf{R}_{xy}$ is the primitive lattice vector with zero *z*-component. The geometry and material phase topology of the unit cell shown in Fig. 2 are again considered for an example problem. The two materials chosen are GaAs (represented in black) and air (represented in white). The ratio of the dielectric constants is 11.4. Figs. 6a and 6b show the photonic band structures for *TE*-polarization and *TM*-polarization, respectively, using the same number of finite elements and the same **k**-space sampling parameters as in Section V A. However, unlike the phononic crystal case above in which the wave field is a vector field, the present models are governed by scalar equations, (25) and (26). The size of the problem in each is therefore smaller, $n = 2025$. The reduced Bloch mode expansion results, which in this example are based on 2-point expansion and use of eight Bloch modes for every selected **k**-point, are in excellent agreement with those of the full model.

## C. Electronic Band Structure Calculations

In this section, we demonstrate the applicability of reduced Bloch mode expansion to electronic band structure calculations for a 3D model. We start with the time-independent single electron Shrödinger equation,

$$-\frac{\hbar^2}{2m}\nabla^2 \psi(\mathbf{k}, \mathbf{x}) + V(\mathbf{x})\psi(\mathbf{k}, \mathbf{x}) = E(\mathbf{k})\psi(\mathbf{k}, \mathbf{x}), \tag{27}$$

where $\hbar$ is the Planck's constant, and *m*, $E(\mathbf{k})$ and $\psi(\mathbf{k}, \mathbf{x})$ are, respectively, the electron effective mass, energy and wave function. The electronic potential $V(\mathbf{x})$ in a perfect crystal satisfies the relation $V(\mathbf{x}) = V(\mathbf{x}+\mathbf{R})$. We consider a 3D generalized Kronig-Penney model potential [53] for a simple cubic lattice,

$$V(\mathbf{x}) = \sum_{i=1}^{3} \overline{V}(x_i) \quad \text{where} \quad \overline{V}(x_i) = \begin{cases} 0, & 0 \leq x_i < a \\ V_0, & a \leq x_i < b \end{cases}. \tag{28}$$

We choose for an example problem the following values for the parameters: $a = 2$ a.u., $b = 3$ a.u. and $V_0 = 6.5$ R$_y$ [26]. The unit cell is discretized into $18 \times 18 \times 18$ uniformly sized 8-node trilinear hexahedral finite elements, i.e., $n_{el} = 5832$. With the application of periodic boundary conditions, the number of degrees of freedom is $n = 5832$ (noting that (27) is a scalar equation). The **k**-space is discretized such that $l_k = 49$, thus a total of $n_k = 193$ **k**-points are evaluated to generate the band structure. Fig. 7 shows the electronic band structure following the **k**-space paths Γ→X→M→R→Γ. As in the previous cases, the reduced Bloch mode expansion results (based here on 2-point expansion and use of eight Bloch modes for every selected **k**-point) provide an excellent approximation to the results of the full model.

## VI. COMPUTATIONAL EFFICIENCY

In addition to the quality of the primary expansion method, a key component in efficient band structure calculations is the technique and algorithm utilized for the solution of the Bloch eigenvalue problem. Several techniques and algorithms have been developed in the literature for fast eigenvalue problem calculations, e.g., [19,24–25], with the computational complexity for formulations yielding sparse Hermitian matrices roughly being around O($n \log n$) for a single **k**-point calculation. For models with many degrees of freedom (large value of $n$), and for calculations with a large number of **k**-points, $n_k$, to be evaluated, the total cost of generating a band structure or density of states plot can be prohibitive. The reduced Bloch mode expansion approach has the advantage that it is compatible with available primary expansion approaches (as demonstrated for real space FE expansion in the above examples) as well as with techniques and algorithms for solving the Bloch eigenvalue problem. Therefore any positive attributes pertaining to employed methods from the point of view of efficiency are in principle still kept and utilized in the reduced Bloch mode expansion model.

A series of calculations were performed on a single computer processor involving 2D plain strain phononic band structure calculations in order to assess the computational efficiency of the reduced Bloch mode expansion approach as a function of the size of the numerical problem. First, a series of calculations with increasing number of finite elements, $n_{el}$, are considered with the value of **k**-space sampling rate fixed to $l_k = 49$ ($n_k = 145$). As a measure of efficiency, the ratio $r$ is defined as

$$r = \frac{\text{band structure calculation time for reduced Bloch mode expansion model}}{\text{band structure calculation time for full model}}. \quad (29)$$

Fig. 8a shows the value of $r$ as a function of $n_{el}$ for both the 2-point and 3-point expansion schemes. It is observed that beyond 500 finite elements the value of $r$ almost levels off at values ranging from 0.05 to 0.15, with the 3-point expansion being slightly more expensive than the 2-point expansion as expected. In Fig 8b, the results for two specific finite element resolutions were considered, a course resolution with $n_{el} = 729$ and a fine resolution with $n_{el} = 3969$. A series of calculations were carried out to show $r$ as a function of $l_k$, again for both the 2-point and 3-point expansion schemes. A significant decrease in the value of $r$ is observed as $l_k$ is increased, and in particular it is shown that an increase in the number of degrees of freedom (as represented by the finite element resolution) adds substantially to the savings brought about by reduced Bloch mode expansion. For the $n_{el} = 3969$ model with 2-point expansion, two orders of magnitude reduction in computational cost is recorded in these calculations, as shown in Fig. 8b. Further reduction in computational cost is expected with more FE model refinement and higher **k**-space sampling rates.

## VIII. SUMMARY AND CONCLUSIONS

Reduced Bloch mode expansion was presented as an approach for efficient and accurate calculation of band structures for periodic media. This modal analysis approach involves expanding the Bloch solution at all calculation **k**-points using, in its discrete form, a selected reduced set of Bloch eigenvectors to form the expansion basis. This basis is selected within the irreducible Brillouin zone at high symmetry points determined by the crystal's structure and group theory, i.e., the Γ, X, M, R points for the 3D simple cubic lattice. Intermediate **k**-points along the lines connecting the high symmetry points can be selected in addition, i.e., a set that would now consist of the Γ, Δ, X, Z, M, T, R, Λ points for the 3D simple cubic lattice. The former selection is referred to as a *2-point expansion*, and the latter as a *3-point expansion*. At each of the reciprocal lattice selection points, a number of Bloch eigenvectors are selected up to the frequency range (or dispersion branch number) of interest for the band structure calculations. Since it is common to initially discretize the periodic unit cell and solution field using some choice of basis, e.g., using finite elements, reduced Bloch mode expansion is practically a secondary expansion that keeps, and builds on, any favorable attributes a primary expansion approach might exhibit.

Results presented for 2D plain strain phononic band structure calculations within and along the boundaries of the IBZ are in excellent agreement with those obtained from the full model. Bloch mode shapes and density of state predictions also agree. While 2-point expansion gives accurate band structures, the slightly more expensive 3-point expansion scheme improves the accuracy further especially at **k**-points far away from the high symmetry points. Two orders of magnitude in reduction of computation time has been recorded for the proposed approach, and more savings are expected for models of larger size and for calculations based on more refined sampling of the reciprocal lattice space (or IBZ circuit lines). The success of the approach was also demonstrated for 2D photonic and 3D electronic band structure calculations.

## APPENDIX: BLOCH ELEMENT

A finite element treatment of (13) yields the algebraic eigenvalue problem given in (20). The global mass and stiffness matrices **M** and **K** are assembled from the local matrices $m^e$ and $k^e$ representing a Bloch element:

$$\mathbf{M} = \underset{e=1}{\overset{n_{el}}{\mathbf{A}}} m^e$$
$$m^e_{pq} = \delta_{ij} \int_{\Omega^e} N_a \rho N_b d\Omega \quad , \tag{A1}$$

where $\delta_{ij}$ is the kronecker delta function defined as

$$\delta_{ij} = \begin{cases} 1, & \text{if } i = j \\ 0, & \text{if } i \neq j \end{cases},$$

and

$$\mathbf{K} = \overset{n_{el}}{\underset{e=1}{\mathbf{A}}} k^e$$
$$k^e_{pq} = \mathbf{e}_i^T \int_\Omega \widetilde{\mathbf{B}}_a^T \mathbf{D} \widetilde{\mathbf{B}}_b d\Omega \mathbf{e}_j \quad , \tag{A2}$$

where $\mathbf{D}$ is the reduced elasticity matrix and $\widetilde{\mathbf{B}}$ is the Bloch-transformed matrix of shape functions and shape function derivatives,

$$\widetilde{\mathbf{B}} = \mathbf{B} + \mathbf{B}^k. \tag{A3}$$

For a number of spatial dimension, $n_{sd} = 2$ (i.e., 2D domain),

$$\mathbf{B}_a = \begin{bmatrix} N_{a,x} & 0 \\ 0 & N_{a,y} \\ N_{a,y} & N_{a,x} \end{bmatrix}, \quad \mathbf{B}_a^k = \begin{bmatrix} ik_x N_a & 0 \\ 0 & ik_y N_a \\ ik_y N_a & ik_x N_a \end{bmatrix}, \tag{A4}$$

and for $n_{sd} = 3$ (i.e., 3D domain),

$$\mathbf{B}_a = \begin{bmatrix} N_{a,x} & 0 & 0 \\ 0 & N_{a,y} & 0 \\ 0 & 0 & N_{a,z} \\ 0 & N_{a,z} & N_{a,y} \\ N_{a,z} & 0 & N_{a,x} \\ N_{a,y} & N_{a,x} & 0 \end{bmatrix}, \quad \mathbf{B}_a^k = \begin{bmatrix} ik_x N_a & 0 & 0 \\ 0 & ik_y N_a & 0 \\ 0 & 0 & ik_z N_a \\ 0 & ik_z N_a & ik_y N_a \\ ik_z N_a & 0 & ik_x N_a \\ ik_y N_a & ik_x N_a & 0 \end{bmatrix}. \tag{A5}$$

**REFERENCES**

[1] A. Messiah, *Quantum Mechanics* (Wiley, New York, 1964).
[2] F. Bloch, Z. Phys. **52**, 555 (1928).
[3] W. Pauli, Z. Phys. **31**, 765 (1925).
[4] R. E. Peierls, Z. Phys. **53**, 255 (1929).
[5] R. E. Peierls, Ann. Phys. **4**, 121 (1930).
[6] A. H. Wilson, Proc. Roy. Soc. London Ser. A **133**, 458 (1931).
[7] A. H. Wilson, Proc. Roy. Soc. London Ser. A **134**, 277 (1931).
[8] E. Yablonovitch, Phys. Rev. Lett. **58**, 2059 (1987).
[9] S. John, Phys. Rev. Lett. **58**, 2486 (1987).
[10] K. M. Ho, C. T. Chan and C. M. Soúkoulis, Phys. Rev. Lett. **65**, 3152 (1990).
[11] M. S. Kushwaha, Int. J. Mod. Phys. B **10**, 977 (1996).
[12] J. D. Joannopoulos, P. R. Villeneuve and S. Fan, Nature **386**, 143 (1997).

[13] M. Sigalas, M. S. Kushwaha, E. N. Economou, M. Kafesaki, I. E. Psarobas, W. Steurer, Zeitschrift Fur Kristallographie **220**, 765 (2005).
[14] D. W. Prather, S. Shi, J. Murakowski, G. J. Schneider, A. Sharkawy, C. Chen and B Miao, IEEE Journal of Selected Topics in Quantum Electronics **12**, 1416 (2006).
[15] L. Brillouin, *Wave Propagation in Periodic Structures* (Dover, New York, 1953).
[16] K. M. Leung and Y. F. Liu, Phys. Rev. Lett. **65**, 2646 (1990).
[17] Z. Zhang and S. Satpathy, Phys. Rev. Lett. **65**, 2650 (1990).
[18] R. D. Meade, A. M. Rappe, K.D. Brommer, J. D. Joannopoulos and O. L. Alerhand, Phys. Rev. B **48**, 8434 (1993).
[19] S. G. Johnson and J. D. Joannopoulos, Optics Express **8**, 173 (2001).
[20] J. B. Pendry and A. MacKinnon, Phys. Rev. Lett. **69**, 2772 (1992).
[21] X. Wang, X. G. Zhang, Q. Yu and B. N. Harmon, Phys. Rev. B **47**, 4161 (1993).
[22] K. M. Leung and Y. Qiu, Phys. Rev. B **48**, 7767 (1993).
[23] H. Y. D. Yang, IEEE Trans. Microwave Theory Tech. **44**, 2688 (1996).
[24] D. C. Dobson, J. Comput. Phys. **149**, 363 (1999).
[25] W. Axmann and P. Kuchment, J. Comput. Phys. **150**, 468 (1999).
[26] J. E. Pask, B. M. Klein, P. A. Sterne and C. Y. Fong, Computer Phys. Communications **135**, 1 (2001).
[27] S. Jun, Y. S. Cho and S. Im, Optics Express **11**, 541 (2003).
[28] E. Moreno, D. Erni and C. Hafner, Phys. Rev. B **65**, 155120 (2002).
[29] X. Checoury and J. M. Lourtioz, Optics Communications **259**, 360 (2006).
[30] Z. Z. Yan and Y. S. Wang, Phys. Rev. B **74**, 224303 (2006).
[31] P. J. Chiang, C. P. Yu and H. C. Chang, Phys. Rev. E **75**, 026703 (2007).
[32] L. C. Botten, N. A. Nicorovici, R. C. McPhedran, C. M. de Sterke and A. A. Asatryan, Phys. Rev. E **64**, 046603 (2001).
[33] M. Marrone, V. F. Rodriguez-Esquerre, H. E. Hernández-Figueroa, Opt. Express **10**, 1299 (2002).
[34] J. Yuan and Y. Y. Lu, J. Opt. Soc. Am. **23**, 3217 (2006).
[35] K. Busch, G. von Freymann, S. Linden, S. F. Mingaleev, L. Tkeshelashvili, M. Wegener, Physical Reports **444**, 101 (2007).
[36] S. J. Cox and D. C. Dobson, SIAM Journal on Applied Mathematics **59**, 2108 (1999).
[37] M. Burger, S. J. Osher and E. Yablonovitch, IEICE Transactions on Electronics **E87C**, 258 (2004).
[38] R. L. Chern, C. C. Chang, C. C. Chang and R. R. Hwang, Phys. Rev. E **68**, 026704 (2003).
[39] T. W. McDevitt, G. M. Hulbert and N. Kikuchi, Comput. Methods Appl. Mech. Eng. 190, 6425 (2001).
[40] G. Nagai, J. Fish and K. Watanabe, Computational Mechanics **33**, 144 (2004).
[41] Z. P. Wang and C. T. Sun, Wave Motion **36**, 473 (2002).
[42] M. I. Hussein and G. M. Hulbert, Finite Elem. Anal. Design **42**, 602 (2006).
[43] S. Gonella and M. Ruzzene, Appl. Mathematical Modelling **32**, 459 (2008).
[44] L. Rayleigh, *The Theory of Sound* (London, 1877).
[45] O. Døssing, Modal Analysis: The International Journal of Analytical and Experimental Modal Analysis 10, 69 (1995).
[46] G. Floquet, Ann. de l'Ecole Normale Supérieur **12**, 47 (1883).

[47] A. Baldereschi, Phys. Rev. B **7**, 5212 (1973).
[48] D. J. Chadi and M. L. Cohen, Phys. Rev. B **7**, 692 (1973).
[49] Z. Y. Li and K. M. Ho, Phys. Rev. B **67**, 165104 (2003).
[50] I. R. Matias, I. Del Villar, F. J. Arregui and R. O. Claus, J. Opt. Soc. Am, A **20**, 644 (2003).
[51] H. T. Zhang, D. S. Wang, M. L. Gong and D. Z. Zhao, Optics Communications **237**, 179 (2004).
[52] T. J. R. Hughes, *The Finite Element Method* (Prentice-Hall, New Jersey, 1987).
[53] R. De L. Kronig and W. G. Penney, Proc. R. Soc. London, Ser. A **130**, 499 (1931).

**Figures**

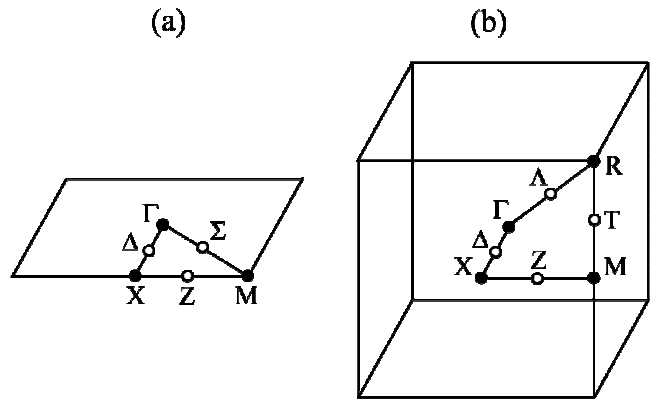

Figure 1. Unit cell in reciprocal lattice space with the irreducible Brillouin zone, high symmetry **k**-points (solid circles) and intermediate **k**-points (hollow circles) shown. (a) 2D square unit cell, (b) 3D simple cubic unit cell.

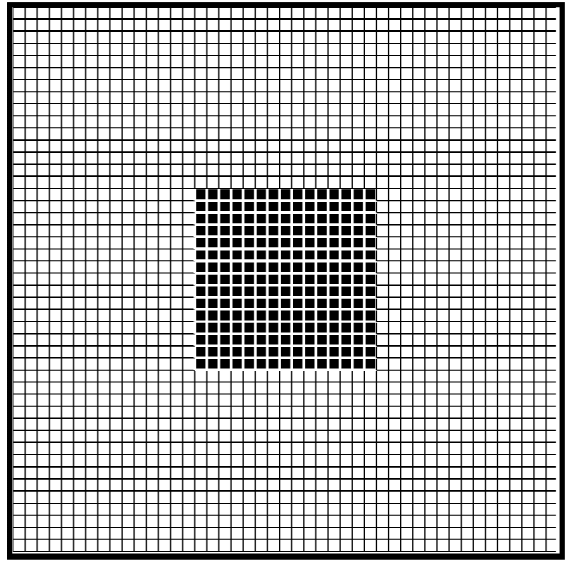

Figure 2. Finite element mesh for 2D square unit cell. The stiff/dense material phase is in black, and the complaint/light material phase is in white.

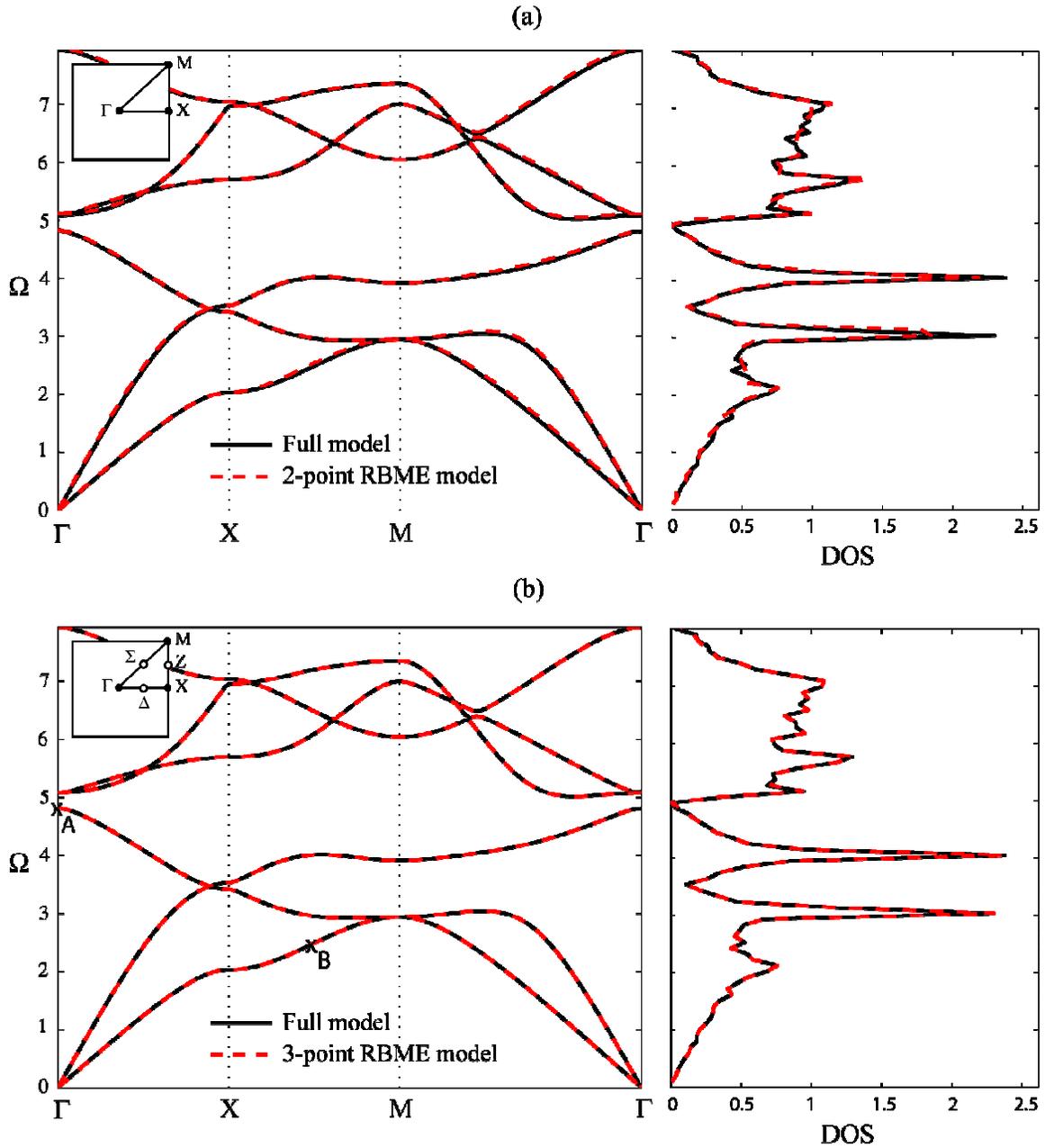

Figure 3. Phononic band structure and density of states (DOS) calculated using full model and reduced Bloch mode expansion model following (a) 2-point expansion, and (b) 3-point expansion. The 2D unit cell, IBZ and eigenvector selection points are shown in the insets.

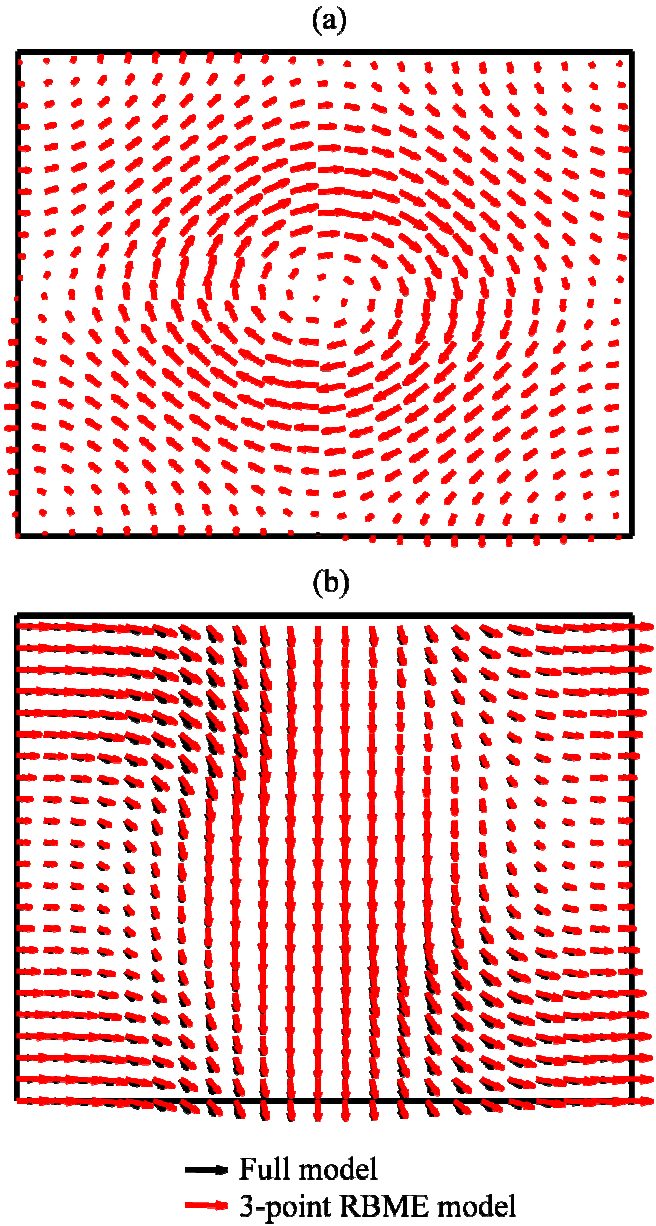

Figure 4. Mode shapes for points A and B in the frequency spectrum as indicated in Fig. 3b. Calculations from the full model and a reduced Bloch mode expansion model following 3-point expansion are shown.

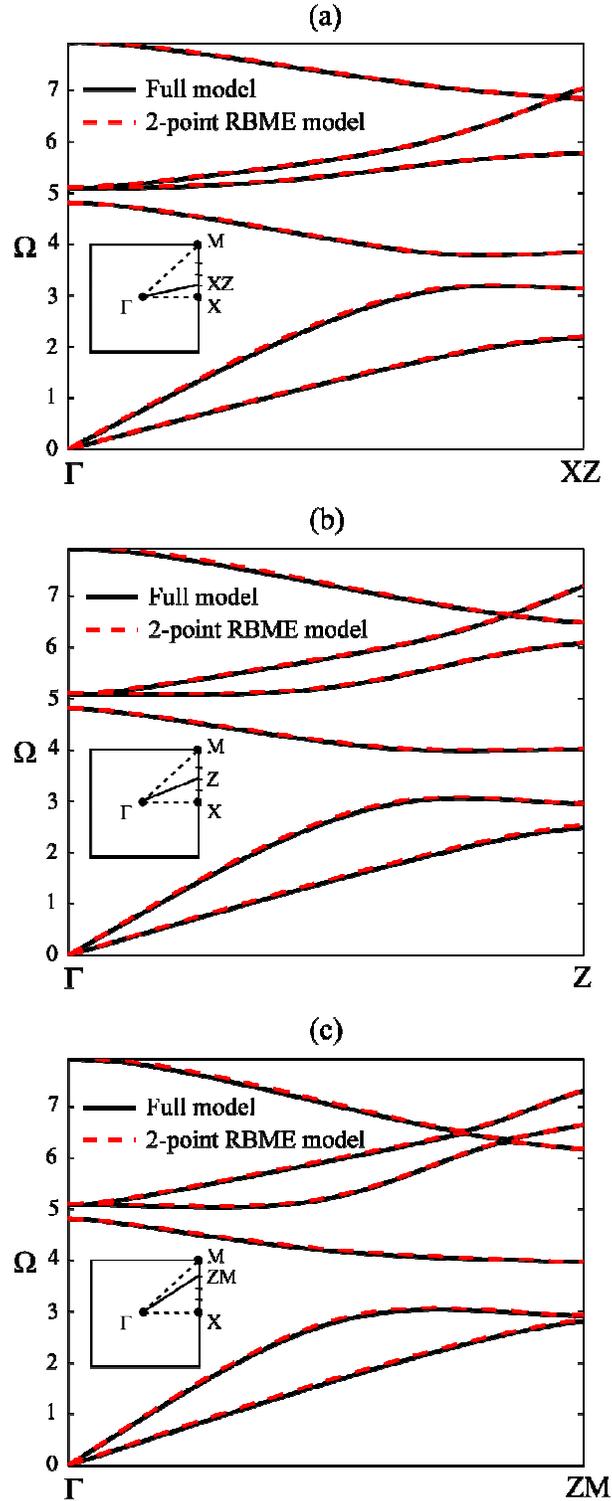

Figure 5. Phononic band structure for different **k**-point sets in the interior of the IBZ. Calculations from the full model and a reduced Bloch mode expansion model following 2-point expansion are shown. The 2D unit cell, IBZ, **k**-point path and eigenvector selection points are shown in the insets.

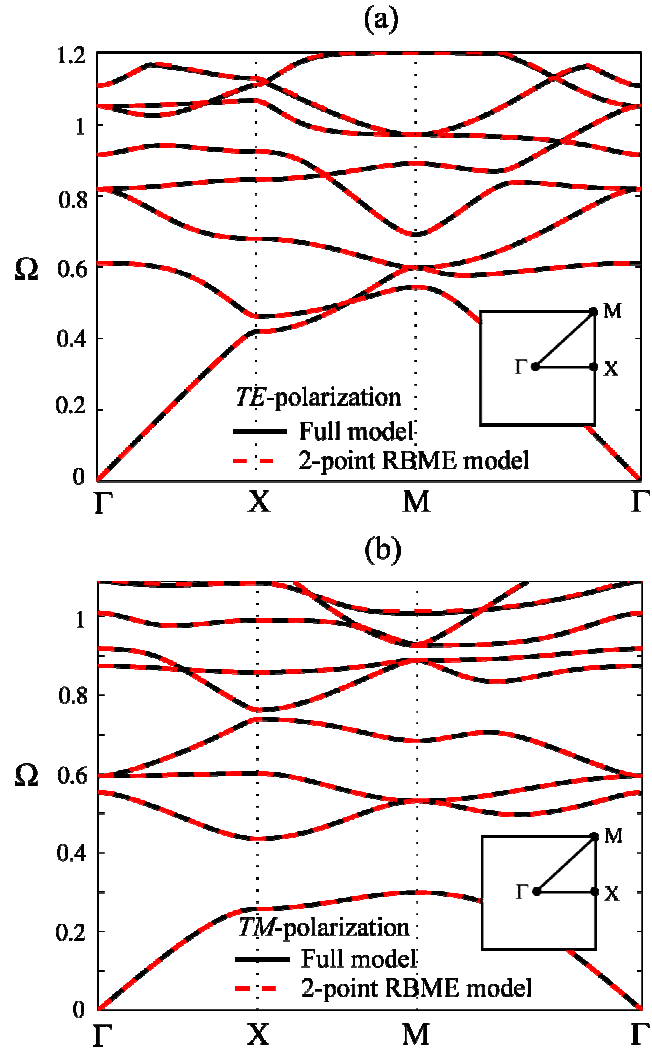

Figure 6. Photonic band structure calculated using full model and reduced Bloch mode expansion model following 2-point expansion. (a) TE polarization, (b) TM polarization. The 2D unit cell, IBZ and eigenvector selection points are shown in the insets.

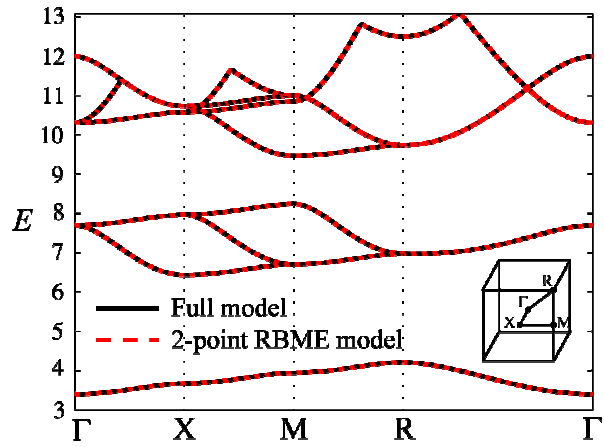

Figure 7. Electronic band structure calculated using full model and reduced Bloch mode expansion model following 2-point expansion. The 3D unit cell, IBZ and eigenvector selection points are shown in the inset.

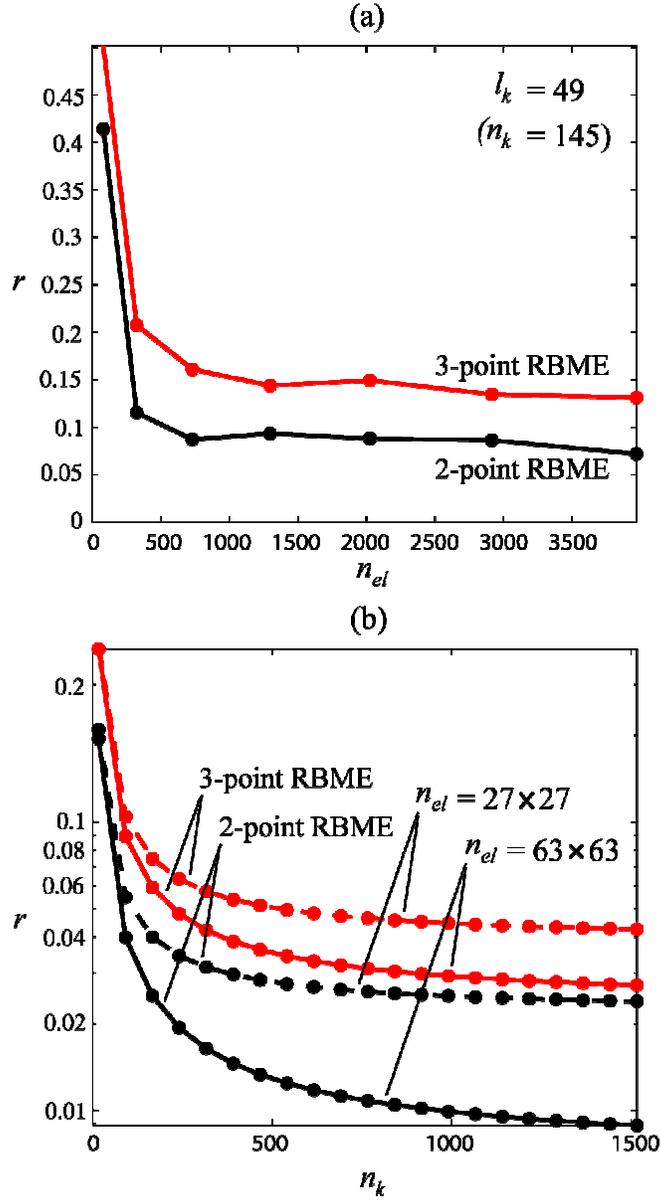

Figure 8. Ratio of reduced Bloch mode expansion model to full model calculation times, $r$, versus (a) number of finite elements, $n_{el}$ (for $l_p = 49$) and (b) number of sampled **k**-points along the border of the IBZ, $n_k$ (for two 2D finite element meshes).